\documentclass[citeautoscript,floatfix,aps,prl,twocolumn,superscriptaddress]{revtex4-2}

\usepackage[english]{babel}
\usepackage{amsmath}
\usepackage{amssymb}
\usepackage{graphicx}
\usepackage[colorlinks=true, allcolors=blue]{hyperref}
\usepackage{tikz}
\usepackage{multirow}
\usepackage{booktabs}
\usepackage{colortbl}
\usepackage{soul}
\usepackage{physics}
\usepackage{verbatim}
\setstcolor{red}
\usepackage{amsmath}

\newcommand{\sro}{Sr$_2$RuO$_4$}{}%
\newcommand{\srho}{Sr$_2$RhO$_4$}{}%
\newcommand{\nco}{NaCo$_2$O$_4$}{}%
\newcommand{\ke}{\mathbf{k}}

\begin{document}

\title{%
Insulating transport in anisotropic metals: 
breakdown of Drude transport and the puzzling $c$-axis resistivity of \sro{} and other layered oxides
}

\author{Sophie  Beck}
\affiliation{Center for Computational Quantum Physics, Flatiron Institute, 162 5th Avenue, New York, New York 10010, USA}
\affiliation{Institute of Solid State Physics, TU Wien, 1040 Vienna, Austria}
\author{Matthew Shammami}
\affiliation{Center for Computational Quantum Physics, Flatiron Institute, 162 5th Avenue, New York, New York 10010, USA}
\affiliation{Department of Chemistry and Biochemistry, University of California, Los Angeles, California 90095, USA}
\affiliation{Mani L. Bhaumik Institute for Theoretical Physics, University of California, Los Angeles, California 90095, USA}
\author{Lorenzo Van Mu\~noz}
\affiliation{Department of Physics, Massachusetts Institute of Technology, 77 Massachusetts Avenue, Cambridge, MA 02139, USA}
\author{Jason Kaye}
\affiliation{Center for Computational Quantum Physics, Flatiron Institute, 162 5th Avenue, New York, New York 10010, USA}
\affiliation{Center for Computational Mathematics, Flatiron Institute, 162 5th Avenue, New York, NY 10010, USA}
\author{Antoine Georges}
\affiliation{Coll\`ege de France, Paris, France and CCQ-Flatiron Institute, New York, NY, USA}
\affiliation{Center for Computational Quantum Physics, Flatiron Institute, 162 5th Avenue, New York, New York 10010, USA}
\affiliation{Centre de Physique Théorique, Ecole Polytechnique, CNRS, Institut Polytechnique de Paris, 91128 Palaiseau Cedex, France}
\author{Jernej Mravlje}
\affiliation{Jožef Stefan Institute, Jamova 39, 1000 Ljubljana, Slovenia}
\affiliation{Faculty of Mathematics and Physics, University of Ljubljana, Slovenia}

\begin{abstract}
  We reveal a mechanism that may explain the non-metallic out-of-plane resistivity in layered metals.
  By carefully examining how  the Drude-Boltzmann expression for the $c$-axis conductivity emerges out of the Kubo formula, we find, besides the standard metallic term proportional to the carrier lifetime $\tau$, a non-Drude contribution proportional to $1/\tau$. 
  The Drude behavior breaks down when $1/\tau > 2 \eta^*$, the crossover value $\eta^*$ being small (and hence observable) when the $c$-axis velocities vary rapidly with the distance from the Fermi surface.
  We consider the Hund metal \sro{} as a test case, which we study within a realistic dynamical mean-field theory approach.
  The non-Drude behavior observed experimentally in $c$-axis transport is reproduced and explained by our considerations, showing that earlier invoked extrinsic mechanisms that involve either impurities or phonons are unnecessary.
  We point out that the small value of $\eta^*$ is due to a peculiar accidental cancellation due to destructive interference characteristic of body-centered tetragonal lattices. 
\end{abstract}

\maketitle

\begin{figure}[h]
  \centering
  \includegraphics[width=\linewidth]{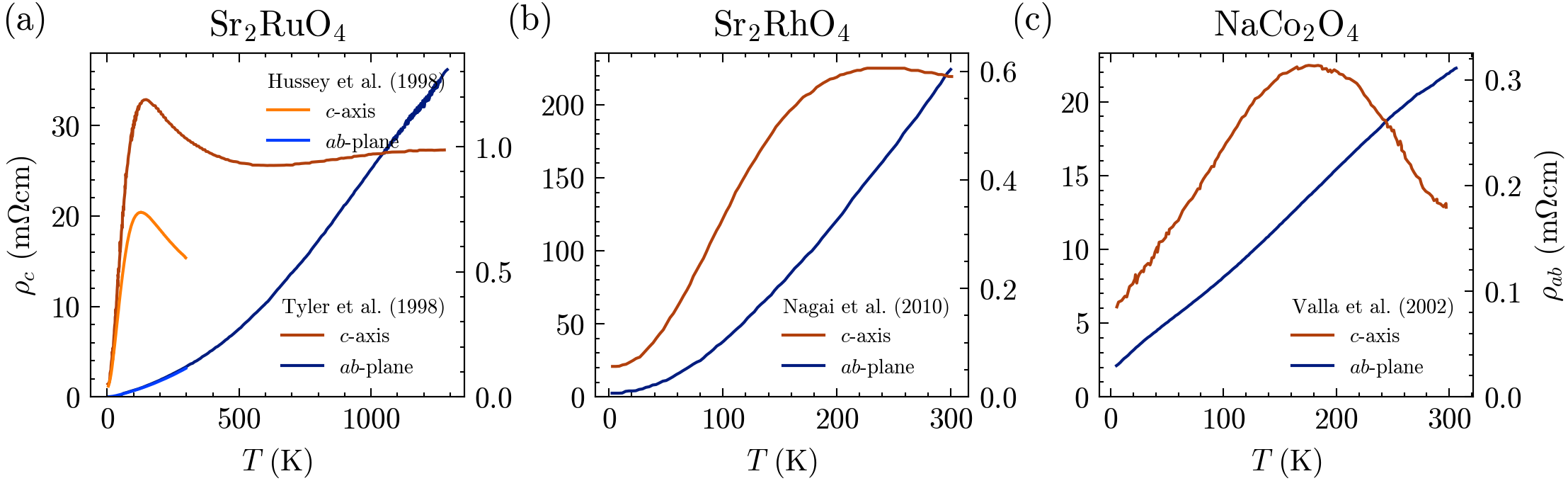}
  \caption{%
  Measurements of in-plane vs. out-of-plane resistivity in layered \sro{} (a), \srho{} (b), and \nco{} (c). The data is digitized from Refs.~\cite{tyler98,Hussey_et_al:1998,Nagai2010,Valla2002}.
  The out-of-plane resistivity exhibits a maximum near the crossover temperature $T_M$, indicating the breakdown of Drude transport.
  }
  \label{fig:exp_overview}
\end{figure}

The electronic response of metals is often described by the simple Drude formula, \mbox{$\sigma = ne^2\tau/m$}, where $n$ is the carrier density, $m$ the carrier mass and $\tau$ the momentum relaxation time.
This expression embodies the intuitive idea that less frequent scattering (i.e., longer $\tau$) enhances electrical conductivity.
Although derived from a classical equation of motion, this framework is used to describe a broad range of metallic systems, including unconventional ones~\cite{Bruin2013,Varma2020,Ataei2022}.
However, this description clearly does not apply in strongly anisotropic systems whenever in- and out-of-plane conductivities behave qualitatively differently.
Notable examples include cobaltates~\cite{terasaki1997} and layered ruthenates~\cite{tyler98}, where the in-plane resistivity remains metallic while the out-of-plane resistivity exhibits insulating-like behavior (see Fig.~\ref{fig:exp_overview}).

These observations have been loosely associated with the ``coherence-incoherence crossover'', i.e. the disappearance of quasiparticles that was reported in photoemission experiments~\cite{Valla2002,wang04} and argued to point to a breakdown of Boltzmann transport.
At first sight this is an appealing picture: when there are no quasiparticles, metallic transport cannot occur.
However, as noted by Millis~\cite{Millis2002}, accepting it raises another question, namely why is then the in-plane transport metallic?

Subsequent work has exposed additional limitations of the coherence-incoherence crossover picture. 
Firstly, recent experiments on \sro{}~\cite{hunter23} have shown that the quasiparticle weight actually does not vanish with increasing temperature and that quasiparticles can be resolved above the crossover temperature $T_M$ associated with the resistivity maximum (see also~\cite{radonjic2010,deng13}). 
Secondly, as first emphasized by Prange and Kadanoff~\cite{prange64}, the existence of long-lived quasiparticles is not a strict prerequisite for the applicability of Boltzmann theory. Boltzmann theory may give accurate results at temperatures far above the Fermi liquid temperatures~\cite{deng13,xu13} or even in the strange metal regime of cuprates~\cite{georges_mravlje_2021,gourgout22}.
Thus, the puzzling question of the $c$-axis resistivity maximum remains open.

This question is especially pertinent in the case of \sro{}, a well-characterized, strongly correlated metal with a body-centered tetragonal structure for which the experiments are particularly controlled due to the existence of large ultra-clean single crystals.
Its electronic properties are determined by four electrons occupying the low-energy $t_{2g}$ orbitals, with the $d_{xz}/d_{yz}$ orbitals defining the out-of-plane conduction.
Transport in this material is highly anisotropic, with $\rho_c/\rho_{ab}$ exceeding 1000 at low temperatures~\cite{tyler98}.
Its Fermi surface and low-energy electronic structure have been accurately captured by density functional theory + dynamical mean-field theory (DFT+DMFT) approaches~\cite{mravlje11,tamai19}, yet an understanding of its transport properties~\cite{abramovitch23}, in particular the origin of the $c$-axis resistivity maximum, remains incomplete.

It should be noted that a substantial body of theoretical work has successfully explained and modeled the $c$-axis maximum through non-electronic, extrinsic mechanisms, invoking either polaronic effects~\cite{lundin03,ho05} or conduction via (resonant) interlayer impurities~\cite{gutman07,gutman08}.
These approaches most often rely on simplified non-interacting models (an important oversimplification, as will become clear in the following) and additional assumptions, while generally not attempting to describe other observables at the same time.
Although the proposed theories are plausible and should be pursued further by carefully assessing their validity through complementary experimental studies, we argue that there is a simpler argument that can equally well explain the anisotropic behavior.
It originates entirely in a peculiarity of the band structure -- destructive interference of hopping amplitudes suppresses contributions to out-of-plane conduction in a way that strongly depends on in-plane momentum -- that leads to a non-isotropic breakdown of the Drude picture.
Namely, as we show in the following, if one starts from band theory and calculates the $c$-axis conductivity from the Kubo formula within a simplification that is valid for strongly anisotropic metals, one finds that the conductivity quite generically behaves as
\begin{equation}
\sigma^c(\tau) = a_0 \tau + a_1 /\tau +  \mathcal{O}(1/\tau^3).
\end{equation}
The second term  dominates when the lifetime becomes sufficiently short:  counterintuitively the conductivity starts to grow with $1/\tau$.


To see this, let  us start from the Kubo formula and let us ignore vertex corrections.
The $c$-axis (DC) conductivity is then given by an expression of the form 
\begin{equation}\label{eq:sigmac}
    \sigma^{c} \propto \int \dd{\omega}  \left(-\frac{\partial f}{\partial \omega}\right) \int \dd[2]{k_{||}} \dd{k}_{\perp} v^z_{\ke{}} A_{\ke{}}(\omega) v^z_{\ke{}} A_{\ke{}}(\omega)\, ,
\end{equation}
where $A_\mathbf{k}(\omega)=(-1/\pi) \mathrm{Im} [\omega +\mu - \epsilon_\mathbf{k} -\Sigma(\omega)]^{-1}$  is the single particle spectral function at energy $\omega$ (given by non-interacting Hamiltonian $\epsilon_\mathbf{k}$, the self-energy $\Sigma$ and chemical potential $\mu$) and $v^z_\mathbf{k}$ is the electron velocity along the $z$-direction.
When applied to the multi-orbital case as in realistic calculations reported later, the quantities become matrices in which case the trace over internal spin/orbital/band indices is implicit.  
We write the momentum integration in terms of the components parallel $k_{||}$ and perpendicular $k_{\perp}$ (labeled as $k$ in the following) with respect to the Fermi surface.

The conductivity is a sum of contributions along the Fermi surface $k_{||}$, i.e. $\sigma^{c}=\int \dd[2]{k_{||}} \sigma^{c}_{k_{||}}$.
To simplify notation, let us consider the case of quasi-1D bands (like the ones that dominate the out-of-plane conductivity of \sro{}), i.e., momentarily ignore any dependence on $k_{||}$. 

To examine how the calculated conductivity varies with $\tau$, let us further assume the spectral functions to be given in terms of a frequency-independent scattering rate $\eta=1/(2 \tau)$ (originating, for instance, from impurity scattering), i.e., they are given by a simple Lorentzian 
\begin{equation}
    A_{k}(\omega) \approx -\frac{1}{\pi} \Im \frac{1}{\omega - v_{\text{F}} k + i \eta}\, ,
\end{equation}
where the band dispersion perpendicular to the Fermi surface has been linearized, $\epsilon_{k} = v_{\text{F}} k$.
At the same time, we retain the momentum dependence in the velocities, and expand in momentum perpendicular to the Fermi surface as
\begin{equation}
  v^z_{k} \approx v_0 +v_1 k +\frac{v_2}{2} k^2.
\end{equation}

We now examine the behavior of $\sigma^{c}$ in the low temperature limit ($T <\eta$), where $-\partial f/\partial \omega \rightarrow \delta(\omega)$ in Eq.~\ref{eq:sigmac}.
Evaluating the remaining integral over momentum we get
\begin{equation}
\label{eq5}
  \sigma^{c} = \frac{\alpha v_0^2}{2 \pi v_F}  \left (\frac{1}{\eta} + \frac{v_1^2 + v_0 v_2}{v_0^2 v_{\text{F}}^2}\eta \right)
\end{equation}
with a unit-dependent constant $\alpha$.
There are, thus, two contributions: a standard metallic Drude term with conductivity proportional to the scattering lifetime $1/\eta$, and a higher-order term directly proportional to the scattering rate $\eta$.
The standard Drude behavior dominates for small $\eta$. 
The corrections come from a change of the velocity away from the Fermi level that leads to an increase of conductivity (if $v_1^2 +v_0 v_2 >0 $, a situation relevant to materials discussed in this paper).
Hence, when $\eta$ becomes larger than a certain limiting value
\begin{equation}\label{eq:etamax}
    \eta^* = \sqrt{\frac{v_{\text{F}}^2 v_0^2}{v_1^2 + v_0 v_2}}\, ,
\end{equation}
the contribution to conductivity at that value of $k_{||}$  goes from a metallic $\propto \tau$ dependence to an insulating regime with $\propto 1/\tau$. 

\begin{figure*}[htbp]
  \centering
  \includegraphics[width=\linewidth]{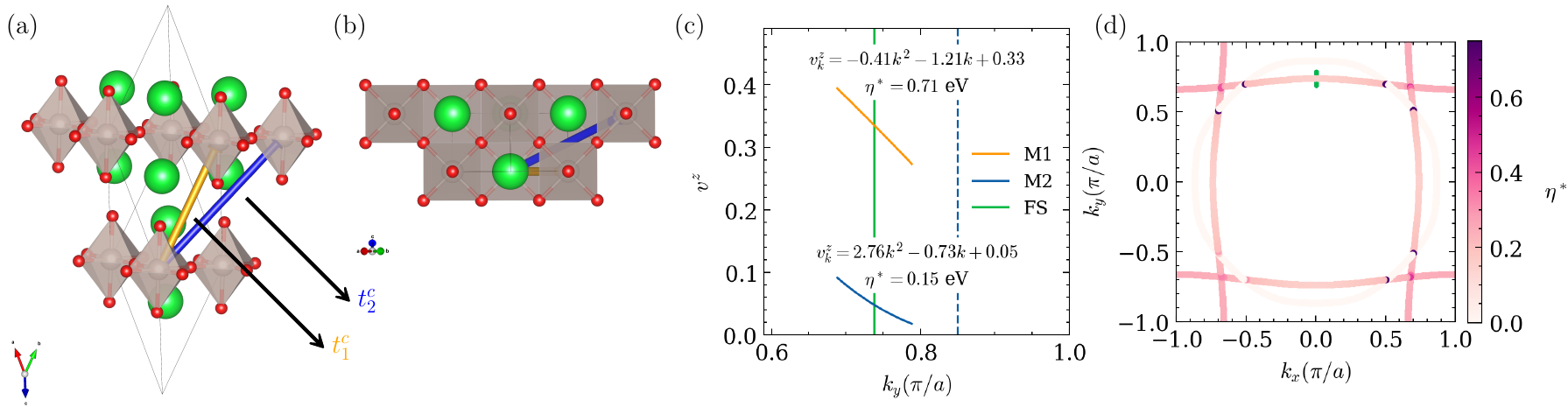}
  \caption{%
  (a) Side view of the body-centered tetragonal lattice of \sro{} indicating the dominant out-of-plane hopping amplitudes.
  (b) Top view.
  (c) The dependence of $v^z$ along a cut perpendicular to the Fermi surface ($k_x = 0.0 \pi/a$) (shown in green in panel d), with expansion in $k$ (with respect to $k_{\text{F}}$) and derived $\eta^*$ values listed, for both the M1 (orange) and M2 model (blue).
  The position of the Fermi surface at $k_{\text{F}}$ is marked in green, the position of the extrapolated vanishing Fermi velocity for M2 is marked in dashed blue.
  (d) The dependence of $\eta^*$ along the Fermi surface.
  }
  \label{fig:eta_on_FS}
\end{figure*}

Several comments are in order: 
(i) Notice that in the derivation we used the peculiarities of the strongly anisotropic band structure.
Retaining the momentum dependence of $v^z$ but truncating the band energy at the linear term $v_{\text{F}} k$ is consistent if the $c$-axis dispersion is negligible compared to the in-plane one. Nonlinearities in the in-plane dispersion on the other hand affect both velocities and the spectral functions; the two effects work in the opposing direction and largely compensate.
This is crucial to understand why the proposed effects apply for the case of the out-of-plane conductivity but {\it not} the in-plane one. 
(ii) Dependence on $k_{||}$ can be restored simply by performing a Fermi surface average of contributions in Eq.~\eqref{eq5}, hence the conclusions are not limited to quasi-1D bands.
(iii) Fine details of the $c$-axis dispersion determine $\eta^*$, and the emergence of non-Drude behavior depends sensitively on those details as will be shown next. 


We now turn to \sro{}, for which the above discussion is directly relevant: the evaluated $\eta^*$ is indeed found to be small enough for the non-Drude contribution to have observable effects in the experimentally accessible temperature range.
We further demonstrate that, in order to properly describe these effects, it is important to retain hopping amplitudes to neighbors beyond the nearest ones in the $c$-direction.
To systematically document this we consider, besides a full ab-initio description of the electronic structure, two simplified models that differ only in the range of $c$-axis hopping included.
The hopping amplitudes are determined from DFT calculations followed by construction of Wannier functions and a careful tight-binding (TB) analysis as outlined in the SM.
The number of in-plane hopping amplitudes does not affect our conclusions; for consistency, we use the six largest ones (see Tab. 1 in the SM). 
For now, we neglect spin–orbit coupling and will return to its effects later.
Along the $c$-axis, only two intra-orbital hopping amplitudes are relevant: the nearest neighbor $t^c_1=-16.2$ meV (orange) and the next-nearest neighbor $t^c_2=-5.9$ meV (blue), as shown in Fig.~\ref{fig:eta_on_FS}.
Model M1 includes the six in-plane hopping amplitudes and $t^c_1$, while Model M2 also includes $t^c_2$.
The corresponding parameters are listed in Tab. 2 in the SM.
We note that truncating hopping amplitudes by size or distance is a common strategy to reduce model complexity, but in this case, such a simplification has drastic consequences.
Remarkably, as shown in the SM the models are virtually indistinguishable if compared on the basis of band structure and Fermi surface alone.
Fig.~\ref{fig:eta_on_FS}c) shows the behavior of $v^z_k$ at a specific point on the Fermi surface, crossing the $\beta$ sheet, comparing the models M1 and M2.
One observes a substantial suppression of the overall magnitude of the velocity (i.e. $v_0$), along with noticeable changes in the quadratic expansion coefficients.
The resulting $\eta^*$ is accordingly reduced by a factor of 5 from 0.71 eV to 0.15 eV going from model M1 to M2 (and is approximately 0.3 eV in the full model, see SM).
The magnitude of $\eta^*$ thus becomes comparable to expected scattering rates around room temperature; the effect is experimentally relevant.
To rule out that this effect is confined to a specific region of the Brillouin zone, Fig.~\ref{fig:eta_on_FS}d) demonstrates that the suppression occurs along the entire Fermi surface.
In the following we illustrate why including $t^c_2$ is essential to reproduce the breakdown of the Drude model as found for the ab-initio model.

\begin{figure}[h]
  \centering
  \includegraphics[width=\linewidth]{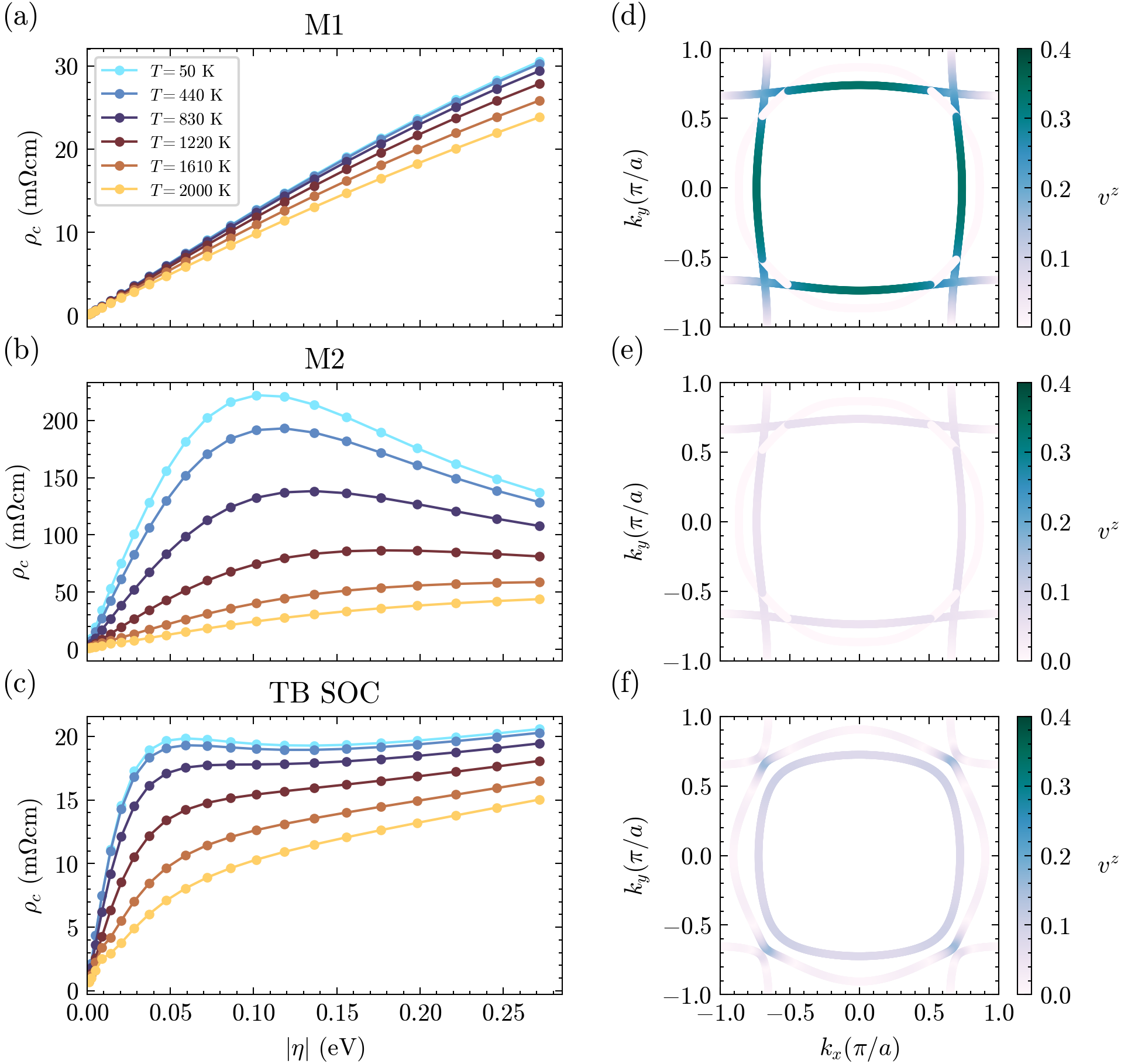}
  \caption{%
  (a-c) Resistivity $\rho^c$ as a function of scattering $\eta$ for the M1 (top), M2 (middle), and full ab-initio model including SOC (bottom).
  (d-f) Corresponding contour maps of $v^z$ at $k_z=0.5 \pi/c$.} 
  \label{fig:rho_vs_eta_tb}
\end{figure}

To investigate the effect on the conductivity, we evaluate the models using the full Kubo formalism while simplifying  scattering to a constant  $\Sigma_m(\omega) \rightarrow -i\eta $, independent of orbital index $m$.
Figure~\ref{fig:rho_vs_eta_tb} shows the resulting resistivities, which show strikingly different behavior.
Model M1 exhibits standard Drude-like behavior, whereas in Model M2 the inclusion of the next-nearest-neighbor hopping causes a pronounced change: the resistivity increases substantially and develops a maximum at $\eta^*$, with a value consistent with the discussion above.
The right panel shows contour plots of the $c$-axis velocity on the Fermi surface, revealing a substantial suppression. This arises from an accidental cancellation due to destructive interference.
Namely, in momentum space, the two dominant $c$-axis hopping amplitudes contribute to the $xz$-$xz$ hopping Hamiltonian as follows (analogously for $yz$-$yz$):
\begin{equation}
\begin{aligned}
  \epsilon_{xz,xz}^{c} = &\left[ t^c_1 \cos(k_x \cdot a/2)  + t^c_2 \cos(k_x \cdot 3a /2) \right]\\
  &\times \cos(k_y \cdot a /2) \cos(k_z \cdot c/2).
\end{aligned}
\end{equation}
Near the Brillouin-zone boundary, $k_x \sim \pi/a$, the contributions to the velocity from these two terms have opposite signs.
Whenever $t^c_2 /t^c_1 > 1/3$ there exists an in-plane momentum $k$ at which the $c$-axis contribution to the hopping vanishes; for the parameters listed in the SM, this occurs at $k = 0.84\pi/a$ (blue dashed line in Fig.~\ref{fig:eta_on_FS}c), i.e. remarkably close to the Fermi momentum (green).
As a result, interference leads to an accidental cancellation of the $c$-axis velocity in close proximity to the Fermi surface.
The Fermi-surface value of the velocity, $v_0$, is thus small, making higher-order corrections comparatively important and leading to a correspondingly small value of $\eta^*$ via Eq.~\ref{eq:etamax}.

An important consequence of the strongly momentum-dependent velocity is a strong temperature dependence of the calculated resistivity.
In Fig.~\ref{fig:rho_vs_eta_tb}a), the averaging around the Fermi surface introduced by the thermal broadening of the Fermi window function, $-\partial f/\partial\omega$, has little effect.
By contrast, in model M2 (panel b), increasing the temperature strongly suppresses the resistivity, as momenta with larger $c$-axis velocity begin to contribute significantly to the $c$-axis response.

These effects survive in the full ab-initio model when all hopping amplitudes and spin-orbit coupling are included (panel c).
The corresponding resistivity is significantly reduced compared to the M2 model, primarily because direct $xz$-$yz$ inter-orbital hopping amplitudes lead to an increase of the conductivity near the Brillouin-zone diagonals.
This contribution softens the maximum in the total resistivity, but the qualitative physics discussed above remains the same (see SM for further details).

\begin{figure}
    \centering
    \includegraphics[width=\linewidth]{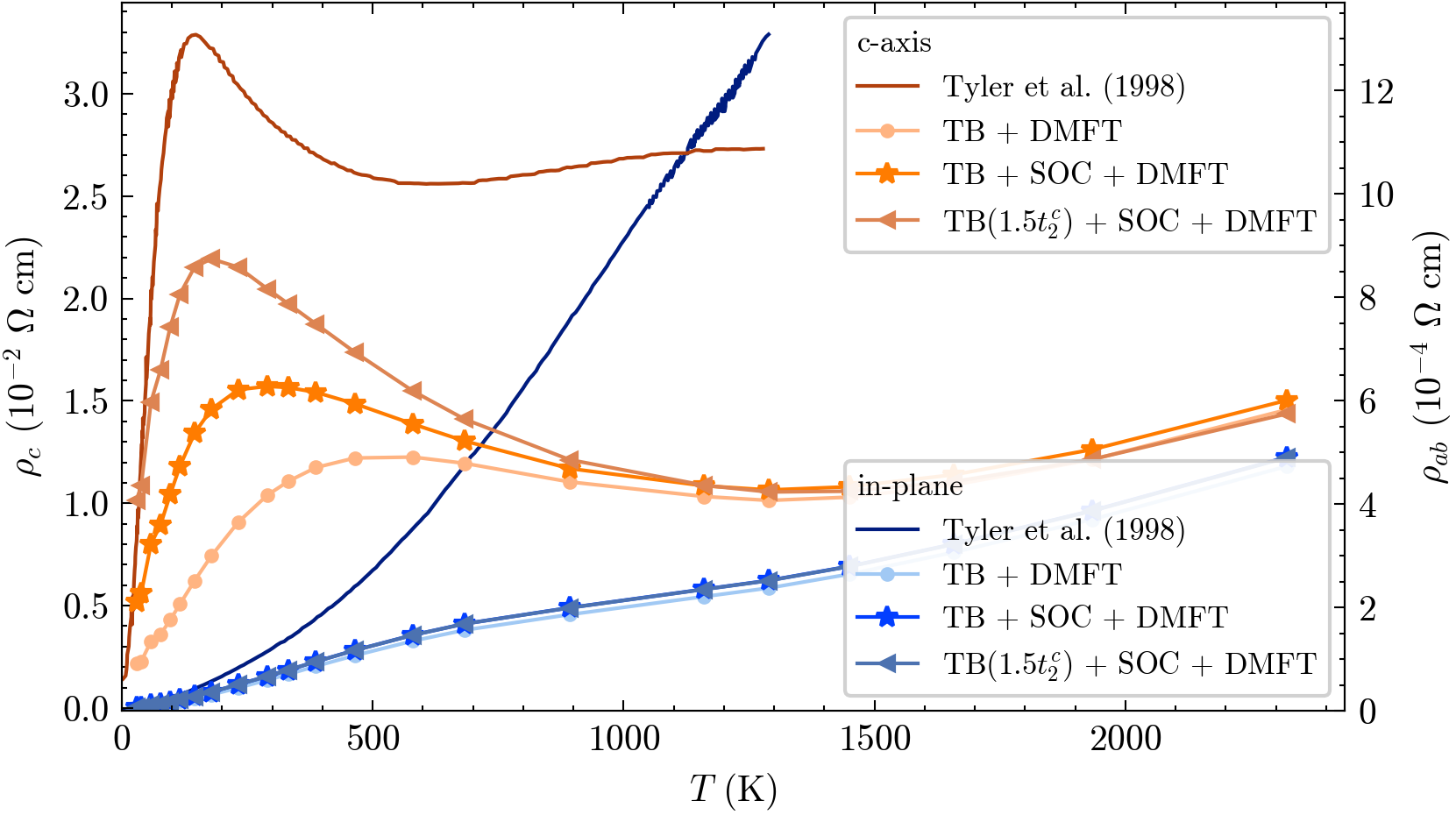}
    \caption{%
    Comparison of out-of-plane resistivity (left axis, orange lines) and in-plane resistivity (right axis, blue lines).
    The $c$-axis resistivity shows a significant dependence on model details, while the in-plane resistivity remains unaffected.
    }
    \label{fig:rho_vs_eta_dmft}
\end{figure}

Finally, we examine electronic many-body effects through the use of orbital-, frequency-, and temperature-dependent DMFT self-energies, following Refs.~\cite{mravlje11,tamai19}, in the spectral functions.
The resulting data, shown in Fig.~\ref{fig:rho_vs_eta_dmft} alongside the digitized experimental results of Ref.~\cite{tyler98}, reveal strongly anisotropic behavior in the full TB+DMFT model: 
the in-plane resistivity (blue) increases monotonically, while the out-of-plane resistivity (orange) shows an intermediate maximum.
The occurrence of the maximum can be traced back to the inclusion of $t^c_2$ that leads to the accidental cancellation, as discussed above and further outlined in the SM.
When spin-orbit coupling is included (TB+SOC+DMFT), the maximum shifts to lower temperature and its peak height increases by about 10\% (consistent with its effect for constant $\eta$, see SM).

As can be seen in Fig.~\ref{fig:eta_on_FS}c), the cancellation is not maximal since $v_0$ is still finite at the Fermi surface (i.e. the green and the orange dashed lines marking the location of the Fermi surface and the extrapolated vanishing Fermi velocity, respectively, do not coincide).
Tuning the region of maximal interference in the Brillouin zone to fall on the Fermi surface by increasing $t^c_2$ by a factor of 1.5 further increases this trend dramatically, producing a sharp peak as is present in the experimental data.
These results demonstrate the extreme sensitivity of the $c$-axis resistivity to very small changes of subdominant hopping matrix elements of the order of only a few meV.

Our theoretical calculations capture some of the salient qualitative aspects of the experiments, making us confident that the proposed mechanism may indeed provide an explanation to the long-standing puzzle of the $c$-axis resistivity maximum in \sro{}.
For instance, it is consistent with the experimental observation that hydrostatic pressure shifts the maximum in $\rho_c$ to higher temperatures while at the same time decreasing the scattering rate~\cite{yoshida1998}.
Chemical doping, inflating the Fermi surface towards the point where accidental cancellation occurs (and increasing the scattering rate), has the opposite effect~\cite{kikugawa2004}.
Our proposed mechanism, which relies on a delicate balance between velocities and scattering rate, naturally explains such a sensitivity to external constraints.
There are however significant differences on the quantitative level, that we now discuss.
Both the in-plane and out-of-plane resistivities are underestimated by theory.
For the in-plane resistivity, this issue was noted earlier in Ref.~\cite{abramovitch23} where it was also shown that a perturbative treatment of the electron-phonon coupling cannot explain the missing scattering.
It is therefore not surprising that the $c$-axis resistivity is likewise underestimated.
Additionally, by comparing with data from Fig.~\ref{fig:rho_vs_eta_tb}, we can infer that an increased magnitude of the scattering would result in a shift of the resistivity maximum to lower temperatures, and could hence improve the agreement with experiment further.  
A possible mechanism for this enhanced scattering is an enhancement of the effective electron-phonon coupling by the strong 
electronic correlations present in this system, for which a proper theoretical formalism needs to be developed~\cite{coulter_2025}.

Does the proposed mechanism apply to other layered materials?
As shown in the SM, data on several other body-centered tetragonal compounds demonstrate that the mechanism indeed applies more broadly.
In particular, our data provides a plausible explanation for the similarly anomalous $c$-axis resistivity maximum in Sr$_2$RhO$_4$~\cite{Nagai2010,Wang_et_al:2024} shown in Fig.~\ref{fig:exp_overview} and predicts plateauing behavior for the $c$-axis resistivity for Sr$_2$FeO$_4$ under pressure but that conversely, in Sr$_2$MoO$_4$ the Drude behavior is obeyed with in-plane and out-of-plane resistivity behaving alike. These predictions should be experimentally tested.
The mechanism likely extends to other lattices, as long as the $c$-axis velocity depends strongly on in-plane momentum, which requires sideways vertical hopping amplitudes comparable to nearest-neighbor direct ones.
Whereas we did not perform specific calculations in that case, we do notice that in layered cobaltates, a strong dependence of the out-of-plane velocity on the in-plane momentum is indeed seen from the results of Ref.~\cite{singh2000}.


In summary, our results explain why Drude behavior can break down in a direction-selective way in anisotropic metals. 
We show that the $c$-axis resistivity becomes non-metallic when the scattering exceeds a characteristic value $\eta^*$, which is small when the electronic velocities vary strongly with momentum perpendicular to the Fermi surface. 
The novel mechanism  is realized in layered ruthenates and rhodates, where $\eta^*$ is small due to a cancellation of out-of-plane velocities resulting from opposite contributions of out-of-plane nearest- and next-nearest-neighbor tunneling paths. 
The mechanism is general and our results call for a reexamination of $c$-axis resistivity in other layered metals, such as cobaltates. 
The existence of a maximum in the $c$-axis resistivity hence does not by itself imply that the transport across layers is incoherent, as often argued.
How the newly revealed mechanism for the insulating transport in the $c$-direction combines with other earlier discussed mechanisms, such as
interplane defects, remains an important question for future work.


\acknowledgements
S.B. acknowledges funding from the Simons Foundation (00010503, AT).
The Flatiron Institute is a division of the Simons Foundation.
J. M acknowledges support by the  Slovenian Research and Innovation Agency (ARIS) under Contract No. P1-0044.

\bibliography{references}
\appendix 

\end{document}


\renewcommand{\thefigure}{S\arabic{figure}}
\renewcommand{\thetable}{S\arabic{table}}
\renewcommand{\theequation}{S\arabic{equation}}

\title{%
Supplemental Material:
Insulating transport in anisotropic metals: 
breakdown of Drude transport and the puzzling $c$-axis resistivity of \sro{} and other layered oxides
}

\author{Sophie  Beck}
\affiliation{Center for Computational Quantum Physics, Flatiron Institute, 162 5th Avenue, New York, New York 10010, USA}
\affiliation{Institute of Solid State Physics, TU Wien, 1040 Vienna, Austria}
\author{Matthew Shammami}
\affiliation{Center for Computational Quantum Physics, Flatiron Institute, 162 5th Avenue, New York, New York 10010, USA}
\affiliation{Department of Chemistry and Biochemistry, University of California, Los Angeles, California 90095, USA}
\affiliation{Mani L. Bhaumik Institute for Theoretical Physics, University of California, Los Angeles, California 90095, USA}
\author{Lorenzo Van Mu\~noz}
\affiliation{Department of Physics, Massachusetts Institute of Technology, 77 Massachusetts Avenue, Cambridge, MA 02139, USA}
\author{Jason Kaye}
\affiliation{Center for Computational Quantum Physics, Flatiron Institute, 162 5th Avenue, New York, New York 10010, USA}
\affiliation{Center for Computational Mathematics, Flatiron Institute, 162 5th Avenue, New York, NY 10010, USA}
\author{Antoine Georges}
\affiliation{Coll\`ege de France, Paris, France and CCQ-Flatiron Institute, New York, NY, USA}
\affiliation{Center for Computational Quantum Physics, Flatiron Institute, 162 5th Avenue, New York, New York 10010, USA}
\affiliation{Centre de Physique Théorique, Ecole Polytechnique, CNRS, Institut Polytechnique de Paris, 91128 Palaiseau Cedex, France}
\author{Jernej Mravlje}
\affiliation{Jožef Stefan Institute, Jamova 39, 1000 Ljubljana, Slovenia}
\affiliation{Faculty of Mathematics and Physics, University of Ljubljana, Slovenia}

\maketitle

\section{Technical details}

\subsection{DFT and TB analysis}

The electronic structure was calculated using the Quantum~ESPRESSO suite~\cite{Giannozzi_et_al:2009} with the PBE functional~\cite{Perdew/Burke/Ernzerhof:1996}.
Within the tetragonal $I4/mmm$ (space group 139) unit cell structural relaxations were performed until all force components were below 0.1mRy/$a_0$ (with $a_0$ the Bohr radius) and stress tensor components below 0.1 kbar, yielding a lattice constant \mbox{$a=7.009$\AA}.
Scalar-relativistic ultrasoft pseudopotentials were employed~\cite{Garrity_et_al:2014}, with plane-wave energy cutoffs set to 60 Ry for the wave functions and 720 Ry for the charge density.
Brillouin-zone sampling was carried out with a $12\times12\times12$ Monkhorst–Pack grid and a Methfessel–Paxton smearing of 0.01 Ry.

The tight-binding Hamiltonian is constructed from the $t_{2g}$ manifold by projecting the ab-initio Hamiltonian onto the three partially-filled ruthenium 4d orbitals using the Wannier90 code~\cite{Pizzi_2020}.
We use the PythTB package~\cite{Coh/Vanderbilt:2022} to analyze and modify the resulting tight-binding Hamiltonian.

\subsection{DMFT and analytic continuation}

The DMFT calculations were performed as described in Ref.~\cite{mravlje11,mravlje16}.
The self-energies were analytically continued to real axis using maximum entropy method~\cite{beach04}.
The interplay between electronic correlations and spin-orbit coupling leads to enhancement of SOC~\cite{kim_2018,tamai19}.
This enhanced value $\lambda=0.2$ eV was used for the calculations that include SOC.

\subsection{Transport calculations}

The transport calculations are performed within the Kubo linear response formalism~\cite{oudovenko06} using the AutoBZ.jl package~\cite{Van-Munoz/Beck/Kaye:2024} providing a modern implementation with efficient numerical integration algorithms~\cite{Kaye2023,VanMunoz_et_al:2024} and interfaces to both the Wannier90 code and the PythTB package.
Convergence was found to be achieved using a $50\times50\times50$ equispaced grid distributed in the full Brillouin zone.


\section{Derivation of Drude breakdown}

At zero temperature for fixed frequency-independent scattering $\eta$ and corresponding spectral functions for the single band case,
\begin{equation} 
A_k(\omega) = -\frac{1}{\pi} \mathrm{Im} \frac{1}{\omega  + \mu - \epsilon_k + i \eta},
\end{equation}
the contribution to $c$-axis conductivity  (Eq.~(2) in the main text) from each 1D slice perpendicular to the Fermi surface  is 
\begin{equation}
    \begin{aligned}
        \sigma^c = \int v_k^2 A_k^2(0) dk &\sim \frac{\eta^2}{\pi^2} \int_{-\infty }^{\infty} (v_k^z)^2  \left[\frac{1}{(v_{\text{F}} k) ^2  + \eta^2}\right]^2  \dd{k}\, \\ 
        &= \frac{\eta^2}{\pi^2} \int_{-\infty }^{\infty} \left[ v_0^2 + (v_1^2 +v_0 v_2) k^2  \right]  \left[\frac{1}{(v_{\text{F}} k) ^2  + \eta^2} \right]^2 \dd{k}\, \\
        &= \frac{1}{2 \pi } \left (\frac{v_0^2}{v_{\text{F}}} \frac{1}{  \eta} + \frac{v_1^2 +v_0 v_2}{v_{\text{F}}^3} \eta \right)\,
    \end{aligned}
\end{equation}
where in  the first step the dependence of band energy on momentum was linearized, $\epsilon_k = v_{\text{F}} k$, and integration domain over momentum variable $k$ was extended to $(-\infty,+\infty)$.
Both these steps are valid if $\eta$ is sufficiently small, such that the contributions to the integral for momenta further from the Fermi surface where the linearization is invalid or boundary of the Brillouin zone is reached give negligible contributions.
This is reasonable because of rapid drop of $A_k^2 \sim 1/k^4$ for $v_{\text{F}} k / \eta \gg 1$.
For the studied regimes a linear approximation of the out-of-plane dispersion turns out to be insufficient.
Hence, we take $v_k^z \approx v_0 + v_1 k +v_2 k^2/2$, i.e. up to second order in the expansion, resulting in the correction term to conductivity that is proportional to $\eta$ (higher orders would lead to even higher powers of $\eta$). 

\section{TB parameters}

\begin{figure}
    \centering
    \includegraphics[width=.45\linewidth]{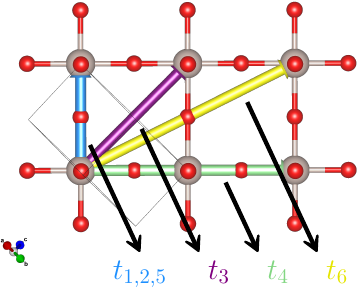}
    \caption{%
    Top view of the 2D Ruthenium plane with the dominant in-plane hopping amplitudes indicated and labeled.
    The corresponding values and orbital characters are given in Tab.~\ref{tab:ip_hoppings}.
    }
    \label{fig:ip_hoppings}
\end{figure}

\begin{figure}
    \centering
    \includegraphics[width=.45\linewidth]{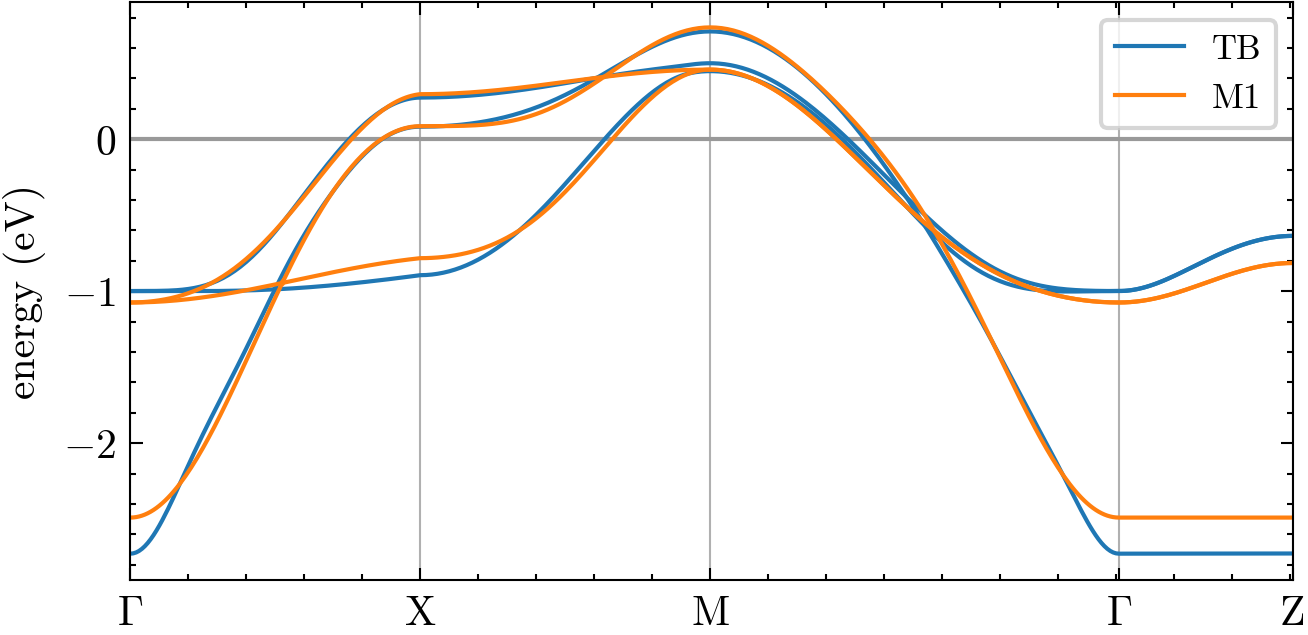}\hspace{10pt}
    \includegraphics[width=.45\linewidth]{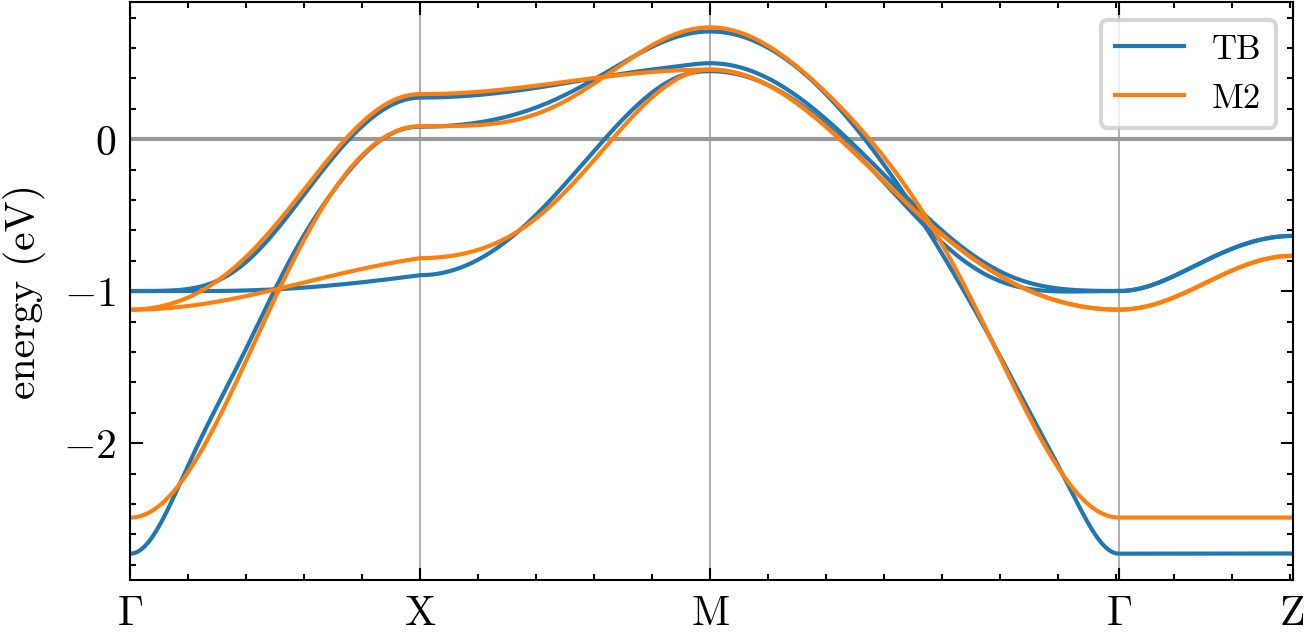}
    \caption{%
    Band structure comparison between the full TB model (blue) and the M1 model (left) and M2 model (right) in orange.
    }
    \label{fig:bands}
\end{figure}

\begin{figure}
    \centering
    \includegraphics[width=.9\linewidth]{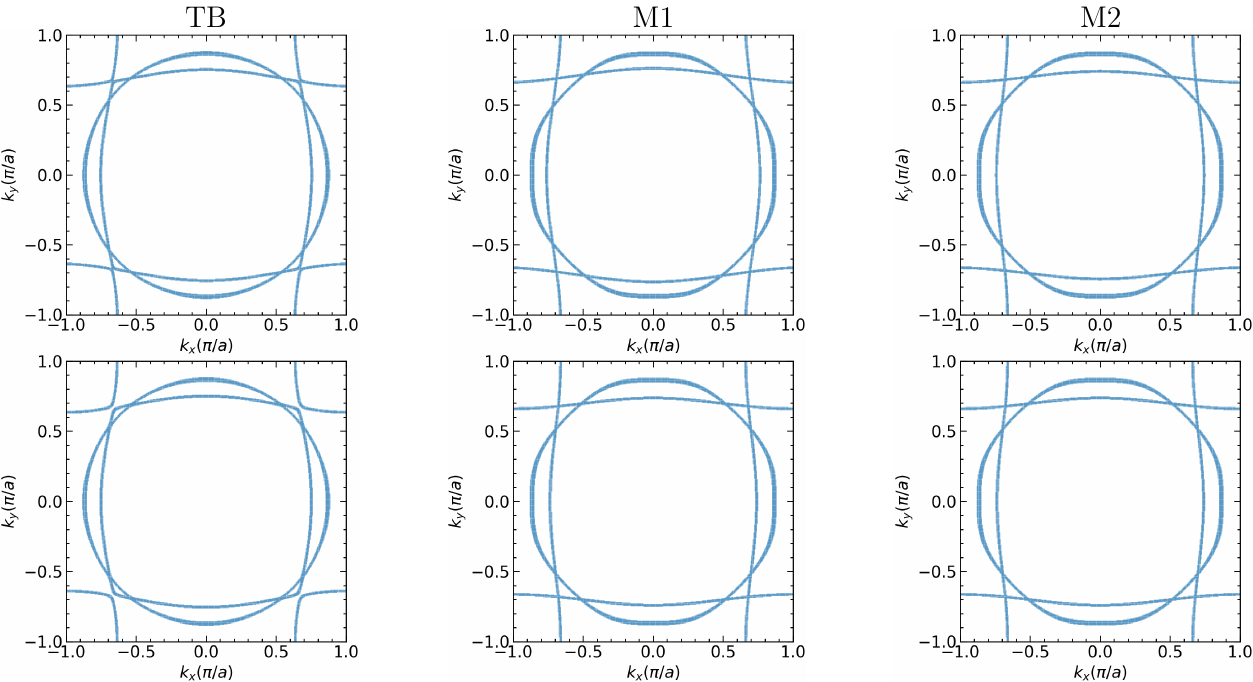}
    \caption{%
    Fermi surface of different models at $k_z=0.0$ (top) and $k_z=0.5$ (bottom) (in units of $\pi/c$).
    The left column shows the data from the full TB model, the middle column from M1, and the right column from M2.
    }
    \label{fig:FS}
\end{figure}

The in- and out-of-plane TB parameters used for the different models (TB, M1 and M2) are given in Tab.~\ref{tab:ip_hoppings} and Tab.~\ref{tab:oop_hoppings}, respectively, with the hopping amplitudes labeled according to Fig.~\ref{fig:ip_hoppings} and Fig. 2a) of the main manuscript.
\begin{table}[h]
\renewcommand{\arraystretch}{1.}
    \centering
    \begin{tabular}{l|cccccc}
        \textbf{Model} & $t_1$ ($xy$) & $t_2$ ($xz$) & $t_3$ ($xy$) & $t_4$ ($xz$) & $t_5$ ($yz$) & $t_6$ ($xy$) \\
        \toprule
        \textbf{TB} & -0.3632 & -0.3102 & -0.1203 & 0.0459 & -0.0404 & -0.02 \\
        \bottomrule
    \end{tabular}
    \caption{%
    Six largest in-plane hoppings used for all models illustrated in Fig.~\ref{fig:ip_hoppings}.
    All hoppings are intra-orbital.
    Orbital characters are indicated in parentheses, where $xz$ also denotes $yz$.
    All values are given in units of eV.    
    }
    \label{tab:ip_hoppings}
\end{table}
\begin{table}[h]
\renewcommand{\arraystretch}{1.}
    \centering
    \begin{tabular}{l|cccc}
        \textbf{Model} & $t^c_1$ ($xz$) & $t^c_2$ ($xz$) & $t^c_{x1}$ ($xz$-$yz$) & $t^c_{x2}$ ($xz$-$yz$) \\
        \toprule
        \textbf{M1} & -0.0162 & 0. & 0. & 0. \\
        \midrule
        \textbf{M2} & -0.0162 & -0.0059 & 0. & 0. \\
        \midrule
        \textbf{TB} & -0.0162 & -0.0059 & $\pm$ 0.0088 & $\pm$ 0.0040 \\
        \bottomrule
    \end{tabular}
    \caption{%
    Largest out-of-plane hoppings for the M1, M2 and full TB model.
    $t^c_{1,2}$ are intra- and $t^c_{x1,2}$ inter-orbital.
    Orbital characters are indicated in parentheses, where $xz$ also denotes $yz$.
    All values are given in units of eV.
    }
    \label{tab:oop_hoppings}
\end{table}
The corresponding band structures and Fermi surfaces are shown in Fig.~\ref{fig:bands} and Fig.~\ref{fig:FS}, respectively.
Based on these quantities alone, there is no significant or even visible difference between the M1 and M2 models.
This is not surprising, since their out-of-plane hopping amplitudes differ by only 6 meV, yet it reinforces the main point of the manuscript: even such small variations can lead to drastic changes in the $c$-axis resistivity.

\section{DMFT self-energy}
\subsection{Temperature dependence}

\begin{figure}
    \centering
    \includegraphics[width=.45\linewidth]{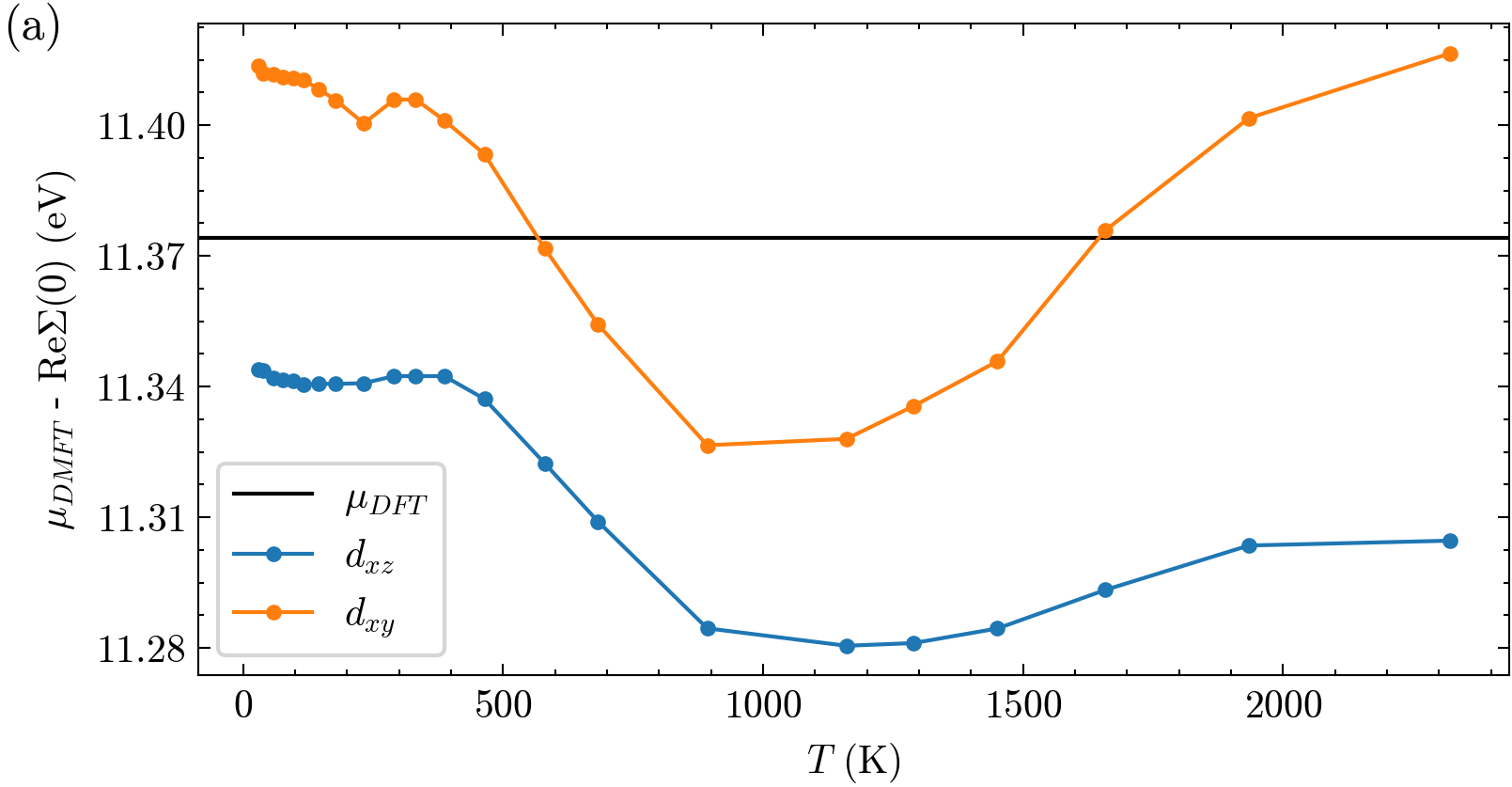}\hspace{10pt}
    \includegraphics[width=.45\linewidth]{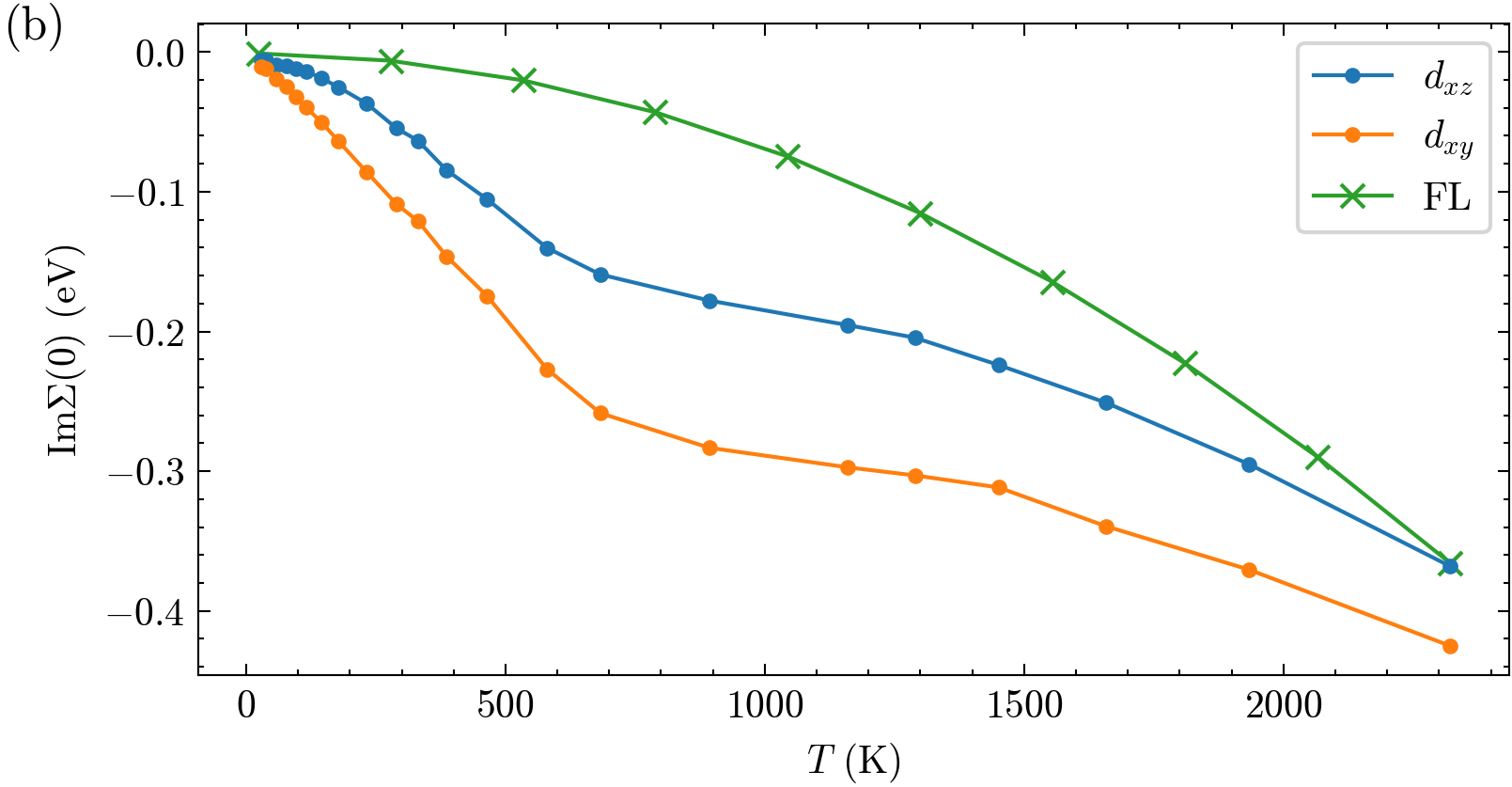}
    \caption{%
    Temperature dependence of DMFT self-energy at $\omega=0$ eV.
    Real part corrected by the DMFT chemical potential is shown in panel (a), the imaginary part in panel (b).
    The green line is a Fermi liquid form $\propto A \pi^2 T^2$ with a prefactor $A$ set so that the magnitude of the self-energy matches the DMFT value at high temperature $\approx 2300$ K.
    }
    \label{fig:DMFT_SE}
\end{figure}

Fig.~\ref{fig:DMFT_SE} shows the effective temperature dependence of the real part of the self-energy $\mu_{\text{DMFT}} - \text{Re}\Sigma(0)$ (panel (a)) and the effective scattering rate $\text{Im}\Sigma(0)$ at zero frequency (panel (b)).
The real part shows a dip at intermediate temperatures between approximately 500-1500 K and then recovers to the low temperature value.
In the same temperature range the scattering rate shows a plateau that sharpens the peak in the response of the $c$-axis resistivity.
The origin of this temperature dependence remains an open question and will be the subject of future studies.

\subsection{Effect of next-nearest neighbor hopping amplitude $t^c_2$}

\begin{figure}
    \centering
    \includegraphics[width=.6\linewidth]{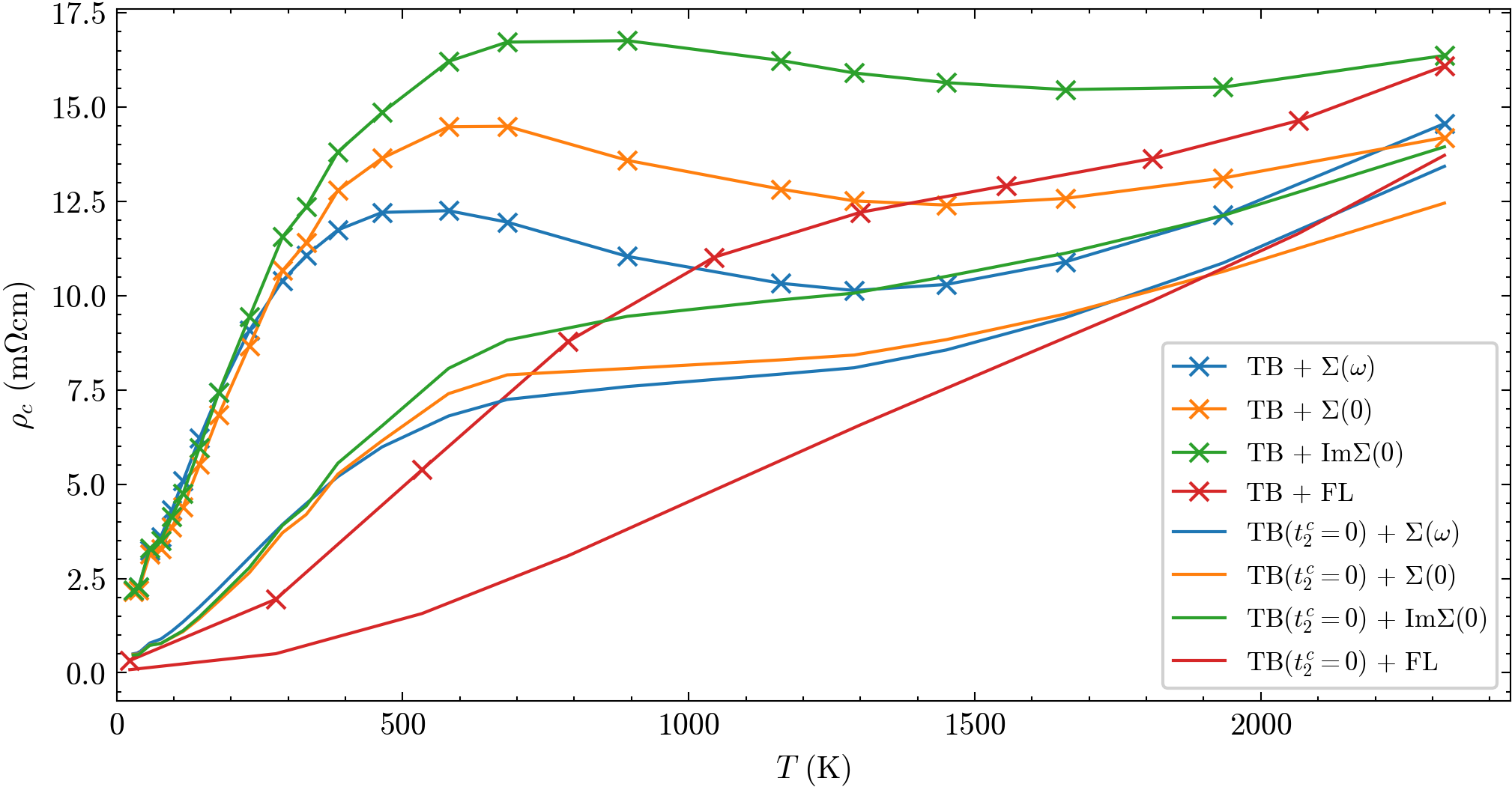}
    \caption{%
    Comparison of the full TB model (lines with crosses) versus the TB model with setting the $t^c_2$ term to zero (solid lines) with various approximations on the DMFT self-energy.
    Blue lines show the full frequency-dependent self-energy (also shown in Fig. 4 in the main manuscript).
    Orange lines correspond to complex self-energies approximated as static using the values at $\omega=0$ eV.
    Blue lines correspond to using only the imaginary part of the self-energies approximated as static scattering rates using the values at $\omega=0$ eV.
    The chemical potential is set to that from DFT.
    Red lines use the FL-approximated self-energy shown in Fig.~\ref{fig:DMFT_SE}b) and the DFT chemical potential.
    }
    \label{fig:rho_comparison}
\end{figure}

To analyze the effect of the frequency and temperature dependence of the DMFT self-energy Fig.~\ref{fig:rho_comparison} shows the $c$-axis resistivity as function of temperature for four different approximations to the self-energy, and for two different TB models.
The blue line with crosses acts as the reference presenting the same data as shown in Fig. 4 of the main manuscript, i.e. the full TB model including the DMFT self-energy.
The orange line with crosses shows the small effect of approximating the full self-energy by its value at $\omega=0$ eV.
For the green line with crosses we further ignore the temperature dependence of the real part of the self-energy shown in Fig.~\ref{fig:DMFT_SE}a) and use the DFT chemical potential.
Finally, the red line with crosses shows an exaggerated Fermi liquid approximation to the DMFT self-energy, i.e. a $T^2$ dependence for all temperatures indicated in green in in Fig.~\ref{fig:DMFT_SE}b).
While this is a purely hypothetical case, it clearly demonstrates a shoulder feature in the resistivity that cannot originate from the temperature dependence of the self-energy, but results from the peculiarity of the TB model.
To demonstrate this, we also show the resistivity of all four discussed cases setting $t^c_2=0$ (removing the interference that leads to the accidental cancellation) as solid lines.
For the FL approximation this eradicates all abnormal temperature dependence of the resistivity, leading to a dominantly $T^2$ behavior.
For the DMFT self-energies, on the other hand, a shoulder feature remains that stems from the temperature dependence of the self-energy.
This demonstrates that for the TB+DMFT data, the accidental cancellation from the TB model and the plateau in the DMFT self-energy act cooperatively.

\section{Other materials}

In this section, we extend the analysis of the main manuscript to three additional materials with a body-centered tetragonal lattice: \srho{}, \sfo{} and \smo{}.
We provide evidence that a similar mechanism is likely at play for \srho{} (as confirmed by experiment), and that it could be detected for \sfo{}, but not for \smo{} if experiments become available.

\subsection{Reference: \sro{}}

\begin{figure}
    \centering
    \includegraphics[width=\linewidth]{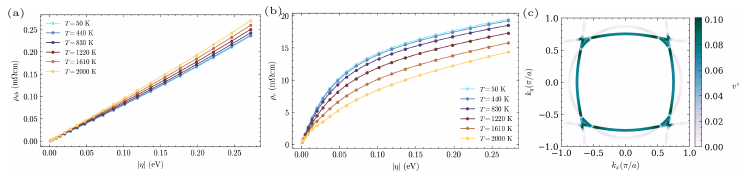}
    \caption{%
    (a) In-plane and (b) out-of-plane resistivity for the ab-initio TB model of \sro{}.
    (c) Contour map of $v^z$ at $k_z=0.5 \pi/c$.
    }
    \label{fig:rho_sro}
\end{figure}

\begin{figure}
    \centering
    \includegraphics[width=\linewidth]{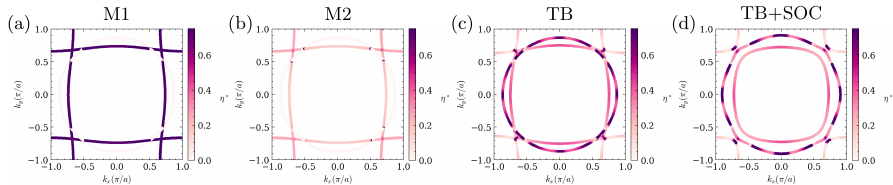}
    \caption{%
    Dependence of $\eta^*$ along the Fermi surface for different models.
    }
    \label{fig:eta_star}
\end{figure}

For the purpose of comparison we first show the data for \sro{} in the absence of SOC in Fig.~\ref{fig:rho_sro}.
Using the full TB model we simulate the resistivity as function of scattering rate.
Figs.~\ref{fig:rho_sro}(a-b) display the in-plane and out-of-plane resistivity, respectively.
While the in-plane resistivity shows the expected $\eta$-linear behavior, the $c$-axis resistivity shows a plateauing behavior with two approximately $\eta$-linear regimes separated by a shoulder at approximately 50 meV and considerable temperature dependence.
The out-of-plane velocities $v^z_k$ shown in Fig.~\ref{fig:rho_sro}c) are suppressed (note the different scale as compared to Fig. 3 in the main manuscript) and remain fairly small in magnitude across the Fermi surface, except at the Brilliouin zone diagonals. 
Fig.~\ref{fig:eta_star} shows the corresponding maps of $\eta^*$ for the different models.
As is clearly visible, adding $t^c_2$ drastically reduces the value along the quasi-1D sheets from M1 to M2.
For the full TB model with and without SOC the inter-orbital hopping terms add weight to the in-plane $d_{xy}$ sheet, but otherwise the picture remains largely unchanged.

\subsection{\srho{}}

\begin{figure}
    \centering
    \includegraphics[width=\linewidth]{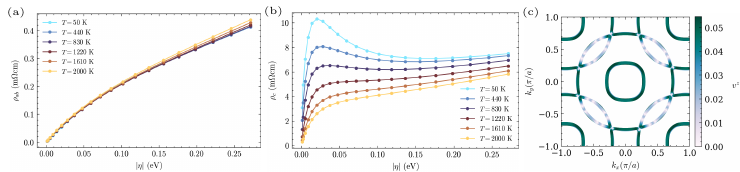}
    \caption{%
    (a) In-plane and (b) out-of-plane resistivity for the ab-initio TB model of \srho{} with $\lambda=0.2$ eV.
    (c) Contour map of $v^z$ at $k_z=0.5 \pi/c$.
    }
    \label{fig:rho_srho}
\end{figure}

For the structure given in Ref.~\cite{srho_data} we construct twelve maximally localized Wannier functions of rhodium 4d character as outlined in the technical details.
Due to the more complex lattice structure with rotations of the four RhO$_6$ octahedra, the TB analysis and subsequent definitions of $t^c_{1,2}$ become considerably more complex and no simple statement can be made as to whether accidental cancellation can occur.
Using the full TB model and an effective spin-orbit coupling $\lambda=0.2$ eV~\cite{Zhang/Pavarini:2019} we thus simulate the resistivity as function of scattering rate.
Figs.~\ref{fig:rho_srho}(a-b) display the in-plane and out-of-plane resistivity, respectively.
While the in-plane resistivity deviates only slightly from the expected $\eta$-linear behavior, the response in the $c$-axis resistivity is even stronger than for \sro{} (compare Fig. 3c).
Accordingly, the out-of-plane velocities $v^z_k$ shown in Fig.~\ref{fig:rho_sfo}c) are strongly suppressed and remain very small in magnitude across the Fermi surface.
These results provide clear evidence for a crossover in the $c$-axis resistivity with a similar mechanism discussed in the main manuscript.
Notably, anomalous $c$-axis resistivity has also been reported experimentally~\cite{Nagai2010,Wang_et_al:2024}.

\subsection{\sfo{}}

\begin{figure}
    \centering
    \includegraphics[width=\linewidth]{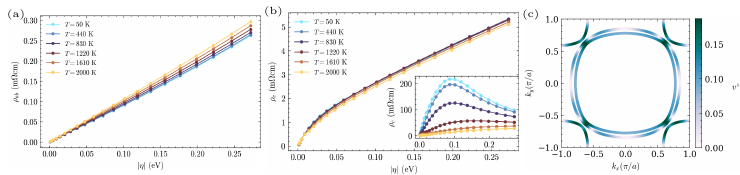}
    \caption{%
    (a) In-plane and (b) out-of-plane resistivity for the ab-initio TB model of \sfo{}.
    (c) Contour map of $v^z$ at $k_z=0.5 \pi/c$.
    Inset in panel (b) shows the out-of-plane resistivity for the corresponding "M2" model.
    }
    \label{fig:rho_sfo}
\end{figure}

In \sfo{} the electronic structure becomes similar to that of \sro{} upon applying 40GPa of isotropic pressure~\cite{Kazemi_et_al:2024}.
Using this structure we construct three maximally localized Wannier functions of iron 3d character as outlined in the technical details.
We repeat the TB analysis and find $t^c_1 = -0.0144$ eV and $t^c_2 = -0.0072$ eV leading to a ratio $r=t^c_1/t^c_2=2<3$, indicating that the velocity suppression mechanism does apply if the location of the Fermi surface is suitably located.
Using the full TB model we calculate the resistivity as function of scattering rate (SOC is small in 3d compounds and taken to vanish in our calculations).
Figs.~\ref{fig:rho_sfo}(a-b) display the in-plane and out-of-plane resistivity, respectively.
While the in-plane resistivity shows the expected $\eta$-linear behavior, the $c$-axis resistivity shows a plateauing behavior with two approximately $\eta$-linear regimes separated by a shoulder at approximately 20 meV.
The inset in Fig.~\ref{fig:rho_sfo}b) shows the results for the analogue of the M2 model for \sfo{} with a clear signature of the resistivity maximum.
The same velocity suppression mechanism
as found for \sro{} is thus at work here, but its effects are less pronounced due to smallness of SOC.
Accordingly, the out-of-plane velocities $v^z_k$ shown in Fig.~\ref{fig:rho_sfo}c) are suppressed and remain fairly small in magnitude across the Fermi surface, except at the Brilliouin zone diagonals. 

We are not aware of experimental results in the literature.
Such measurements would be highly valuable, as our calculations predict a plateau-like behavior.

\subsection{\smo{}}

\begin{figure}
    \centering
    \includegraphics[width=\linewidth]{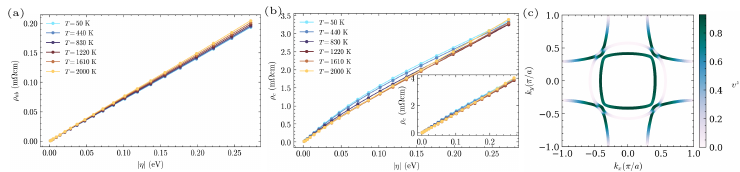}
    \caption{%
    (a) In-plane and (b) out-of-plane resistivity for the ab-initio TB model of \smo{} with $\lambda=0.1$ eV.
    (c) Contour map of $v^z$ at $k_z=0.5 \pi/c$.
    Inset in panel (b) shows the out-of-plane resistivity for the corresponding "M2" model.
    }
    \label{fig:rho_smo}
\end{figure}

\smo{} has been pointed out as the particle-hole dual to \sro{}~\cite{Karp_et_al:2020}.
For the structure given in Ref.~\cite{smo_data} we construct three maximally localized Wannier functions of molybdenum 4d character as outlined in the technical details.
We repeat the TB analysis and find $t^c_1 = -0.0241$ eV and $t^c_2 = -0.0012$ eV leading to a ratio $r=t^c_1/t^c_2=20>3$, indicating that the velocity suppression mechanism is unlikely to apply here.
Using the full TB model and a spin-orbit coupling $\lambda=0.1$ eV~\cite{Karp_et_al:2020} we simulate the resistivity as function of scattering rate.
Figs.~\ref{fig:rho_smo}(a-b) display the in-plane and out-of-plane resistivity, respectively.
Both the in-plane and out-of-plane resistivity show the expected $\eta$-linear behavior, with no effect from removing the next-nearest neighbor hopping amplitude $t^c_2$ in the analogue of the M2 model shown in the inset in Fig.~\ref{fig:rho_smo}b).
Currently, there are no experimental results available to compare against, but our results indicate that there is no signature of the velocity suppression mechanism.
We predict Drude behavior with similar temperature dependence of the $c$-axis and the in-plane resistivity.
Accordingly, the out-of-plane velocities $v^z_k$ shown in Fig.~\ref{fig:rho_smo}c) is large in magnitude across the Fermi surface.

\subsection{Cobaltates}

We evaluate $c$-axis velocity from band dispersions given in Ref.~\cite{singh2000}.
These show strong dependence of the bilayer splitting (that is proportional to $c$-axis velocity) on inplane momentum.
From the data we evaluated $\eta^* \sim 0.1$eV, comparable to what we found for \sro{}.
This tells that an interference between the hoppings to nearest neighbors and  the further neighbors is important also in cobaltates and that it may play a role for the observed $c$-axis maximum also there.
Material-specific calculations should be performed in future work. 

\bibliography{references}